\documentstyle[12pt,graphics]{article}

\input{tcilatex}

\begin{document}

\bigskip

{\bf Langevin processes, agent models and socio-economic systems}

Peter Richmond and Lorenzo Sabatelli, Department of Physics, Trinity
College, Dublin 2, Ireland

\section*{Abstract}

We review some approaches to the understanding of fluctuations in some models used to describe socio and economic systems. Our approach builds on the development of a simple Langevin equation that characterises stochastic processes. This provides a unifying approach that allows first a straightforward description of the early
approaches of Bachelier. We generalise the approach to stochastic equations
that model interacting agents. Using a simple change of variable, we
show that the peer pressure model of Marsilli and the wealth dynamics model
of Solomon are closely related. The methods are further shown to be
consistent with a global free energy functional that invokes an entropy term
based on the Boltzmann formula.

A more recent approach by Michael and Johnson maximised a Tsallis entropy
function subject to simple constraints. We show how this approach can be developed from an agent model where the simple Langevin process is now conditioned by local rather than global noise. The approach yields a BBGKY type hierarchy of equations for the system correlation functions. Of especial interest is that the results can be
obtained from a new free energy functional similar to that mentioned above
except that a Tsallis like entropy term replaces the Boltzmann entropy term.
A mean field approximation yields the results of Michael and Johnson.

We show how personal income data for Brazil, the US, Germany and the UK,
analysed recently by Borgas can be qualitatively understood by this approach.

\section*{Agent models}

We consider a simple type of agent model consisting of a set \{i) of agents,
each of which can have an intrinsic attribute, {\it x}$_{i}$. This could be
a measure of attractiveness or repulsiveness. It could be proportional to
the financial health or wealth of the agent. Now assume that this gives rise
to some kind of peer pressure or a force that tends to make `people keep up
with the Jones' as they say in the UK. Such an interaction force {\it f}(%
{\it x}$_{i}${\it -x}$_{j}$) between the different agents is very similar to
that between molecules interact ing via an inter-particle potential. The
strength of this kind of `social' force, as has been pointed out by for
example, Krawiecki, Holyst and Helbing \footnote{%
A Krawiecki, J A Holyst and D Helbing, arXiv: cond-mat/0210044 v1 2 Oct 2002}
, is not dependent on spatial neighbourhood; rather its origin lies in
communication between agents. The spatial topology of agents may be
considered to be unimportant. The simplest type of peer pressure is
introduced by assuming

\begin{equation}  \label{eq1}
f\left( {X_{i} - X_{j}} \right) = - f\left( {X_{j} - X_{i}} \right)
\end{equation}

Contrast this with the usual assumption for molecular systems where ${\it f}(%
{\it x})={\it f}(|{\it x}|)$. The peer pressure force has not only a
magnitude but also a direction depending on whether the argument, $x$, is
positive or negative.

Now assume the agent dynamics can be captured by the following set of
Langevin equations:

\begin{equation}  \label{eq2}
\frac{{dX_{i}} }{{dt}} = \xi _{i} + \sum\limits_{j} {f\left( {X_{i} - X_{j} }
\right)}
\end{equation}

The stochastic terms, $\{ \xi _{i} \} $ are defined as:

\begin{equation}  \label{eq3}
\left\langle {\xi _{i} \left( {t} \right)\xi _{j} \left( {t^{\prime}} \right)%
} \right\rangle = 2D\delta _{ij} \delta \left( {t - t^{\prime}}
\right);\quad \quad \quad \left\langle {\xi _{i} \left( {t} \right)}
\right\rangle = 0
\end{equation}

We now compute the many agent probability distribution function

\begin{equation}  \label{eq4}
P_{N} \left( {x,t} \right) = \left\langle {\prod\limits_{i = 1}^{N} {\delta %
\left[ {X_{i} \left( {t} \right) - x_{i}} \right]}} \right\rangle
\end{equation}

A standard calculation gives:

\begin{equation}  \label{eq5}
\frac{{\partial P_{N} \left( {x|t} \right)}}{{\partial t}} = \frac{{D}}{{2}}%
\sum\limits_{i} {\frac{{\partial ^{2}P_{N} \left( {x|t} \right)}}{{\partial
x_{i}^{2}} }} - \sum\limits_{i,j} {\frac{{\partial }}{{\partial x_{i}} }} %
\left[ {f\left( {x_{i} - x_{j}} \right)P_{N} \left( {x|t} \right)} \right]
\end{equation}

Reducing the N coordinate representation yields a hierarchy of equations of
agent probability distribution functions:

\bigskip

\begin{equation}  \label{eq6}
\frac{{\partial P_{1} \left( {x|t} \right)}}{{\partial t}} = \frac{{D}}{{2}}%
\frac{{\partial ^{2}P_{1} \left( {x|t} \right)}}{{\partial x^{2}}} - \frac{{%
\partial} }{{\partial x}}\int {dx^{\prime}f\left( {x - x^{\prime}}
\right)P_{2} \left( {x,x^{\prime}|t} \right)}
\end{equation}

This hierarchy will be recognised as a dynamical generalisation of the BBGKY
hierarchy used in the theory of liquids. Clearly a solution is only possible
if we have a route to truncating this infinite hierarchy of coupled
equations. The simplest way forward is to assume a mean field approximation:

\begin{equation}  \label{eq7}
P_{2} \left( {x,x^{\prime}|t} \right) \to P_{1} \left( {x|t} \right)P_{1}
\left( {x^{\prime}|t} \right)
\end{equation}

Dropping now the suffix when N=1 we obtain

\begin{equation}  \label{eq8}
\frac{{\partial P\left( {x|t} \right)}}{{\partial t}} = \frac{{D}}{{2}}\frac{%
{\partial ^{2}P\left( {x|t} \right)}}{{\partial x^{2}}} - \frac{{\partial} }{%
{\partial x}}\phi \left( {x|t} \right)P\left( {x|t} \right)
\end{equation}

The mean field force is given by:

\begin{equation}  \label{eq9}
\phi \left( {x|t} \right) = \int {dx^{\prime}f\left( {x - x^{\prime}}
\right)P\left( {x^{\prime}|t} \right)}
\end{equation}

Now consider the agent pair force {\it f.} We have noted above (equation (%
\ref{eq1}) that it has a directional character. Clearly a number of
functions fit this form. A particularly simple choice is:

\begin{equation}
\begin{array}{l}
f\left( {x}\right) =ae^{-x/\zeta }\quad \quad x>0 \\ 
f\left( {x}\right) =-ae^{x/\zeta _{1}}\quad \ \ x<0
\end{array}
\label{eq10}
\end{equation}

If $\alpha $ \TEXTsymbol{>} 0, we see that agents with attribute ${\it x}%
_{j} $ \TEXTsymbol{<} ${\it x}_{i}$ are attracted to agent {\it i}. Equally
agents with attribute {\it x}$_{j}$ \TEXTsymbol{>} {\it x}$_{i}$ choose to
avoid agent {\it i.} This mimics a kind of `keeping up with Jones' social
force.

This now leads to the expression:

\begin{equation}  \label{eq11}
\phi \left( {x|t} \right) = a\int\limits_{x}^{\infty} {dx^{\prime}e^{ -
\left( {x^{\prime}- x} \right)/\zeta} P\left( {x^{\prime}|t} \right)} -
a\int\limits_{ - \infty} ^{x} {dx^{\prime}e^{ - \left( {x - x^{\prime}}
\right)/\zeta _{1}} P\left( {x^{\prime}|t} \right)}
\end{equation}

This is almost, but not quite, equivalent to the form used by Marsili 
\footnote{%
F Cecconi, M Marsili, JR Banavar \& A Maritan Diffusion, peer pressure and
tailed distributions asXiv: cond-mat/0202212v2 2 Jul 2002} who used
essentially a normalised form of the above. Our form will prove useful in
illustrating the link to the Lotka Volterra approach developed by Solomon
and co-workers\footnote{%
S Solomon and P Richmond Power Laws of Wealth, market order and market
returns Physica A {\bf 299} 2001 188-197} .

\subsection*{Generalised Lotka Volterra Equations - approach of Solomon}

Using our particular choice of force defined in equation (\ref{eq10}), we
introduce the change of variable

\begin{equation}  \label{eq12}
w_{i} = e^{X_{i} /\zeta}
\end{equation}

Now use the theorem of Ito \footnote{%
See for example Financial Calculus: An Introduction to Derivative Pricing, M
Baxter \& A Rennie, Cambridge University Press 1996} , which states that if
X is a stochastic process satisfying:

\begin{equation}  \label{eq13}
dX_{t} = \sigma _{t} dZ_{t} + \mu _{t} dt
\end{equation}

Where {\it Z}$_{t} $ is a Wiener or Brownian process, then providing {\it %
F(x)} is a continuous, twice differentiable function {\it Y}$_{t}${\it =F}(%
{\it X}$_{t}$) is also a stochastic process and is given by

\begin{equation}  \label{eq14}
dY_{t} = \left( {\sigma _{t} F^{\prime}\left( {X_{t}} \right)} \right)dZ_{t}
+ \left( {\mu _{t} F^{\prime}\left( {X_{t}} \right) + \frac{{\sigma _{t}^{2} 
}}{{2}}F^{\prime\prime}\left( {X_{t}} \right)} \right)dt
\end{equation}

Thus the stochastic equation for our new variable {\it w}$_{i}$ defined in (%
\ref{eq12}) is:

\begin{equation}  \label{eq15}
\zeta \frac{{dw_{i}} }{{dt}} = w_{i} \varepsilon \left( {t} \right) +
a\sum\limits_{w_{j} < w_{i}} {w_{j}} - a\sum\limits_{w_{j} > w_{i}} {\frac{{%
w_{_{i}} ^{1 + \zeta /\zeta _{1}} } }{{w_{_{j}} ^{\zeta /\zeta _{1} }} }} + 
\frac{{D}}{{2}}w_{i}
\end{equation}

Note that now the stochastic function $\varepsilon _{i} \left( {t} \right)$
takes only positive values and the additional term D may be absorbed within
the mean value of this stochastic function.

We refer to (\ref{eq15}) as GLVPP (Generalised Lotka Volterra with Peer
Pressure). Note that they are exact consequences of introducing the specific
force (\ref{eq10}) into the initial Langevin dynamics defined by equations (%
\ref{eq2}). No mean field approximation has been used to reach this point.

For large values of $w_{i} $, the third term on the RHS of equation (\ref
{eq15}) may be neglected. Equally the sum over $w_{j} < w_{i} $ may be
extended over all $w_{j} $. As a result equation (\ref{eq15}) reduces
essentially to the generalised Lotka-Volterra equations proposed by Solomon
for wealth dynamics. These latter equations have been solved using both a
mean field approach and exact numerical methods. Both approaches show that
the distribution of wealth, {\it w,} displays a power law or fat tail 
\footnote{%
O Malcai, O Biham P Richmond and S Solomon, Phys Rev {\bf E 66} 2002 31-102}
.

The comparison between our simple peer pressure model and the GLV model of
Solomon is shown in Figure 1. The points are computed from the Langevin
equation not the mean field theory so is an exact solution. Clearly the
difference for the parameters used is minimal.

The equation does not exhibit time correlations that will lead to clustered
volatility. Solomon and Louzoun \footnote{%
Y Louzoun \& S Solomon Volatility driven market in a generalised Lotka
Volterra formalism arXiv: cond-mat/0109493 v1 26 Sept 2001} have however
generalised the equation essentially by replacing the random term, $%
\varepsilon \left( {t} \right)$with the non-linear random term that includes
a `dynamic volatility' and proceed to show that this modification leads to
time auto correlation functions for the volatility that exhibit slow long
time decay in line with observations. In some respects this is similar to
that of the Heston model

\subsection*{The Heston model}

Dragulescu and Yakovenko have studied a particular stochastic model termed
the Heston model \footnote{%
A D Dragulescu and V M Yakovenko arXiv: cond-mat/0203046 v3 5 Nov 2002} .
This is characterised by two equations:

\begin{equation}  \label{eq16}
\frac{{dw}}{{dt}} = fw + D\left( {t} \right)^{1/2}w\varepsilon ^{1}\left( {t}
\right)
\end{equation}

\begin{equation}  \label{eq17}
\frac{{dD}}{{dt}} = - \gamma \left( {D\left( {t} \right) - \bar {D}} \right)
+ \kappa D\left( {t} \right)^{1/2}\varepsilon ^{2}\left( {t} \right)
\end{equation}

The volatility, D(t), is here assumed to fluctuate about the mean, $\bar {D}$
with characteristic relaxation time $1/\gamma $. A so-called variance noise $%
\kappa $is also introduced. The second stochastic process, $\varepsilon
^{2}\left( {t} \right)$ is assumed to be correlated with the first i.e. $<
\varepsilon ^{1}\varepsilon ^{2} > = \rho $ where -1< \~{n}< 1.

Known in the finance literature as the Cox-Ingersoll-Ross process, and in
the mathematical literature as an example of a Feller process, equations (%
\ref{eq16}) and (\ref{eq17}) can be solved completely in terms of modified
Bessel functions. Dragulescu and Yakovenko show that their solutions fit
well Dow Jones data compiled over the period 1882 -2001 and 1990 - 2001. The
data scales over time differences for the returns of 10 to 250 days for
particular values of the parameters $f,\quad \gamma ,\quad \bar {D}\quad
and\quad \kappa $. The authors suggest that it will fit other financial time
series. It remains to study the time autocorrelation functions.

The approach has also been studied in a rather different way by Kozuki \&
Fuchikami \footnote{%
N Kozuki and N Fuchikama arXiv: cond-mat/0210090 v1 4 Oct 2002} . These
authors effectively solve equation (\ref{eq16}) for the stationary
conditional probability {\it p}({\it w}, {\it D}). Having also obtained the
stationary solution to equation (\ref{eq17}), they use standard relations to
obtain the unconditional probability distribution, {\it p}({\it w}) which of
course agrees with the stationary solution obtained by Dragulescu and
Yakovenko.

This is an interesting approach based on empirical data for D(t) that
suggests it may be fitted by an equation of the form (\ref{eq17}). It
extends the work of Bachelier but the underlying dynamics in terms of
agents, as has been also noted by Johnson et al \footnote{%
NF Johnson, P Jefferies and PMH Hui Financial Market Complexity, Oxford
University Press, Oxford, 2003} , remains unexplained. Even more recently,
Dover \footnote{%
Y Dover preprint submitted to Physica A} has shown how using a fluctuating
entropy function that is then averaged over an ensemble of entropy sub
systems, the Tsallis entropy also emerges.

\section*{Entropy functional approach}

It is often possible for physical systems to recover the Fokker-Planck
dynamics from consideration of a free energy functional for the system.
Rather than go directly to expressions involving averages, we can go by way
of Ito's theorem. Return to the Langevin equation (\ref{eq2}) and consider
the density function$\rho _{i} \left( {x,t} \right) = \delta \left( {x -
X_{i} \left( {t} \right)} \right)$. Applying Ito's theorem yields the
stochastic differential equation:

\begin{equation}
\frac{\partial \rho _{i}}{\partial t}=\frac{\partial }{\partial X}(\xi
_{i}(t)\rho _{i}(x,t)+\frac{\partial }{\partial X}(\int dyf(x-y)\rho
(y,t)\rho _{i}(x,t)+D\frac{\partial ^{2}\rho _{i}}{\partial x^{2}}
\label{eq18}
\end{equation}

Now exploit the fact that the Dirac delta function satisfies the property: $%
\frac{\partial {{\delta \left( {x-X}\right) }}}{\partial X}{=}\frac{\partial 
{{\delta \left( {x-X}\right) }}}{\partial x}$and we see that equation (\ref
{eq18}) reduces to equation (\ref{eq6}) as it should when we average over
the random variables. Using this property and summing over the agent index 
{\it i} gives a stochastic differential equation for the density $\rho
\left( {x,t}\right) =\sum\limits_{i}{\delta \left( {x-X_{i}\left( {t}\right) 
}\right) }$:

This is not yet a closed equation for the total density function $\rho
\left( {x,t} \right)$. However, we now follow Dean \footnote{%
D S Dean J Phys A Math Gen {\bf 29} No 24 21 December 1996 L613-617} and
introduce the noise term

This has the correlation function:

\begin{equation}  \label{eq21}
\left\langle {\xi \left( {x,t} \right)\xi \left( {x^{\prime},t^{\prime}}
\right)} \right\rangle = 2D\delta \left( {t - t^{\prime}} \right)\sum%
\limits_{i} {\frac{{\partial} }{{\partial x}}} \frac{{\partial} }{{\partial
x^{\prime}}}\rho _{i} \left( {x,t} \right)\rho _{i} \left( {%
x^{\prime},t^{\prime}} \right)
\end{equation}

Again using the property noted above of delta functions, this expression
simplifies to:

\begin{equation}  \label{eq22}
\left\langle {\xi \left( {x,t} \right)\xi \left( {x^{\prime},t^{\prime}}
\right)} \right\rangle = 2D\delta \left( {t - t^{\prime}} \right)\frac{{%
\partial} }{{\partial x}}\frac{{\partial} }{{\partial x^{\prime}}}\delta
\left( {x - x^{\prime}} \right)\rho \left( {x,t} \right)
\end{equation}

We may now redefine the noise field in terms of a global noise field:

\begin{equation}  \label{eq23}
\xi ^{\prime}\left( {x,t} \right) = \frac{{\partial \eta \left( {x,t}
\right)\rho ^{1/2}\left( {x,t} \right)}}{{\partial x}}
\end{equation}

The white noise field, $\eta $, satisfies:

\begin{equation}  \label{eq24}
\left\langle {\eta \left( {x,t} \right)\eta \left( {x^{\prime},t^{\prime}}
\right)} \right\rangle = 2D\delta \left( {x - x^{\prime}} \right)\delta
\left( {t - t^{\prime}} \right);\quad \quad \left\langle {\eta \left( {x,t}
\right)} \right\rangle = 0
\end{equation}

The Gaussian noises \^{i} and \^{i}' have the same correlation functions and
are statistically identical. Equation (\ref{eq19}) may then be re-expressed
as a closed Langevin equation for the global density of the system:

\begin{equation}  \label{eq25}
\frac{{\partial \rho} }{{\partial t}} = D\frac{{\partial ^{2}\rho \left( {x,t%
} \right)}}{{\partial x^{2}}} - \frac{{\partial} }{{\partial x}}\left( {\eta
\left( {x,t} \right)\rho ^{1/2}\left( {x,t} \right)} \right) - \frac{{%
\partial} }{{\partial x}}\left( {\int {dyf\left( {x - y} \right)\rho \left( {%
y,t} \right)\rho \left( {x,t} \right)}} \right)
\end{equation}
Consider now the following {\bf {\it coarse-grained}} free energy functional

\begin{equation}  \label{eq26}
F\left[ {\rho} \right] = \int {dx\int {dx^{\prime}} \rho \left( {x}
\right)V\left( {x - x^{\prime}} \right)} \rho \left( {x^{\prime}} \right) +
D\int {dx\rho \left( {x} \right)log\rho \left( {x} \right)}
\end{equation}

The potential {\it V} is defined in terms of the force in the usual way, i.e.%
$f\equiv -{\partial V/\partial x}$. The Langevin equation (\ref{eq25}) for
the density, \~{n}, is recovered now using the following prescription:

\begin{equation}
\frac{{\partial \rho }}{{\partial t}}=\frac{{\partial }}{{\partial x}}\left( 
{\rho \frac{{\partial }}{{\partial x}}{\frac{{\delta F}}{{\delta \rho }}|}%
_{\rho =\rho \left( {x,t}\right) }}\right) +\xi ^{\prime }\left( {x,t}\right)
\label{eq27}
\end{equation}

Where the global noise field \^{i}' is given by equation (\ref{eq23}).

\section*{`Tsallis' processes}

Tsallis and colleagues have proposed the following so called entropy
function \footnote{%
CJ Tsallis Stat Phys {\bf 52} 1988 479} :

\begin{equation}  \label{eq28}
S_{q} = - \frac{{1}}{{1 - q}}\left( {1 - \int {dxP^{q}\left( {x,t} \right)} }
\right)
\end{equation}

It is easy to see and has been pointed many times that this expression
reduces to the Boltzmann entropy in the limit that q tends to zero although
the fundamental basis for the postulated form of Tsallis is contentious.
Nevertheless it has been used to describe a wide range of phenomena
associated with criticality and complex systems. In particular Michael and
Johnson have shown how this may be used to fit S\&P stock market data 
\footnote{%
F Michael and MD Johnson arXiv: cond-mat/0108017 v1 1 Aug 2001} . They
maximise the Tsallis entropy given by equation (\ref{eq28}) together with
the following constraints:

\begin{equation}  \label{eq29}
\int {P\left( {x,t} \right)dx} = 1
\end{equation}

\begin{equation}  \label{eq30}
\left\langle {x - \bar {x}\left( {t} \right)} \right\rangle _{q} \equiv \int 
{\left( {x - \bar {x}} \right)} P^{q}\left( {x,t} \right)dx
\end{equation}

\begin{equation}  \label{eq31}
\left\langle {\left( {x - \bar {x}\left( {t} \right)} \right)^{2}}
\right\rangle _{q} \equiv \int {\left( {x - \bar {x}} \right)^{2}}
P^{q}\left( {x,t} \right)dx = \sigma _{q}^{2} \left( {t} \right)
\end{equation}

This yields

\begin{equation}  \label{eq32}
P\left( {x,t} \right) = \frac{{1}}{{Z\left( {t} \right)}}\left\{ {1 + \beta
\left( {t} \right)\left( {q - 1} \right)\left[ {x - \bar {x}\left( {t}
\right)} \right]^{2}} \right\}^{ - 1/\left( {q - 1} \right)}
\end{equation}

The Lagrange multipliers Z and \^{a} are given by

\begin{equation}  \label{eq33}
Z\left( {t} \right) = \frac{{B\left( {\frac{{1}}{{2}},\frac{{1}}{{q - 1}} - 
\frac{{1}}{{2}}} \right)}}{\sqrt {\left( {q - 1} \right)\beta \left( {t}
\right)} }
\end{equation}

\begin{equation}  \label{eq34}
\beta \left( {t} \right) = \frac{{1}}{{2\sigma _{_{q}} ^{2} \left( {t}
\right)Z^{q - 1}\left( {t} \right)}}
\end{equation}

B(x,y) = $\Gamma $(x)$\Gamma $(y)/$\Gamma $(x+y) is Euler's Beta function
and the usual variance of the distribution is

\begin{equation}
\sigma ^{2}=\left\langle {\left( {x-\bar{x}}\right) ^{2}}\right\rangle
_{q=1}=\int {dx\left( {x-\bar{x}}\right) ^{2}}P\left( {x,t}\right) ={\ 
\begin{array}{l}
{\frac{{1}}{{\left( {5-3q}\right) \beta }},\quad q<5/3} \\ 
{\infty ,\quad \quad \ \ \ \ \ q\geq 5/3}
\end{array}
}  \label{eq35}
\end{equation}

Clearly for application to data with finite variance, the parameter q must
lie within the range 1 to 5/3.

This approach seems to yield some of the important stylised features of
stock data, power laws for example. Does this approach have a link to
Langevin processes and can we obtain a free energy functional in the manner
of the previous section that describes the system?

Let's return to our liquid analogue introduced at the beginning of this
paper. In the simple approach of Einstein, it is assumed that `collisions'
of our colloid particle are only with the molecules that comprise the
supporting fluid. However imagine that the colloid particle is now actually
one of the molecules of the fluid. So we are considering collisions between
the fluid molecules themselves. Each molecule still undergoes random
collisions however it now seems plausible to assume that the strength of
each collision is proportional to the density of the fluid. The simplest
approximation is to assume an average density. However suppose that our
fluid is undergoing strong density fluctuations of the kind one might expect
close to a critical point. Now we might assume that the amplitude of our
random collision process is proportional not to the global density but to
some power, {\it \'{i}}, of the local density, {\it \~{n}}({\it x,t}).

With the addition of a drift force {\it f(x)}, the Langevin equation now
takes the following form:

\begin{equation}
\begin{array}{l}
\frac{{dX_{i}}}{{dt}}=f\left( {X}\right) +\xi _{i}\left( {t}\right) \rho
^{\nu }\left( {x,t}\right) \\ 
where\quad as\quad before \\ 
\rho \left( {x,t}\right) \equiv \sum\limits_{j}{\delta \left( {x-X_{j}\left( 
{t}\right) }\right) }
\end{array}
\label{eq36}
\end{equation}

The idea of using such a construction may seem a little strange when
thinking about liquid state theory. It will only be even approximately valid
if the local density fluctuation has a life time much greater than the
collision process and this aspect may require more investigation. Certainly
the process is open to criticism. However when thinking about agent models,
such an interaction seems a plausible construction for the manner in which
agents interact with others. The idea of local noise is a familiar one
amongst analysts who estimate earnings and value of stock prices. Local
consensus frequently occurs! Bear in mind that the variable, x now is not a
spatial parameter. Rather it is an attribute of the agent. One might then
follow the results discussed earlier and choose a different form for the
amplitude. Perhaps agents receive information not only from the `local'
community but also agents holding greater wealth. In this case one might
choose a term of the kind $\int_{x}^{\infty} {\rho ^{\mu} \left( {x,t}
\right)dx} $. Here we shall stick with equation (\ref{eq36}) that at least
has the merit, as we shall see, of leading to a particularly interesting
solution.

Using Ito's theorem, we can now deduce the dynamic equation for the density $%
{\it \rho }_{i}$:

\begin{equation}  \label{eq37}
\frac{{\partial \rho _{i} \left( {X\left( {t} \right)} \right)}}{{\partial t}%
} = - \frac{{\partial f\left( {x} \right)\rho _{i}} }{{\partial x}} - \frac{{%
\partial} }{{\partial x}}\left( {\xi _{i} \left( {t} \right)\rho ^{\nu}
\left( {x,t} \right)\rho _{i} \left( {x,t} \right)} \right) + D\frac{{%
\partial ^{2}}}{{\partial x^{2}}}\left( {\rho ^{2\nu} \left( {x,t}
\right)\rho _{i} \left( {x,t} \right)} \right)
\end{equation}

Taking the average to compute the probability distribution and summing each
side over {\it I} and using the fact that the density function is a delta
function yields:

\begin{equation}
\frac{{\partial P\left( {x,t}\right) }}{{\partial t}}=-\frac{{\partial {\it f%
}\left( {x}\right) P\left( {x,t}\right) }}{{\partial x}}+D\frac{{\partial
^{2}P_{2\nu +1}\left( {x,t}\right) }}{{\partial x^{2}}}  \label{eq38}
\end{equation}

Once more we have a hierarchy of coupled equations. Now we may apply mean
field theory to the second term on the RHS of (\ref{eq38}):

\begin{equation}  \label{eq39}
P_{2\nu + 1} \to \left( {P} \right)^{2\nu + 1}
\end{equation}

Introducing {\it q} = 2{\it \'{i}}+1 yields

\begin{equation}  \label{eq40}
\frac{{\partial P\left( {x,t} \right)}}{{\partial t}} = - \frac{{\partial
\left( {f\left( {x} \right)P\left( {x,t} \right)} \right)}}{{\partial x}} + D%
\frac{{\partial ^{2}P^{q}\left( {x,t} \right)}}{{\partial x^{2}}}
\end{equation}

Choosing the force {\it f} $\sim $ {\it A}+{\it Bx} it is readily seen that
the stationary solution to this equation takes the form given by equation (%
\ref{eq32}). This suggests that we might be able to formulate a free energy
functional that involves the Tsallis formula. Summing equation (\ref{eq37})
over the index {\it i} gives

\begin{equation}
\frac{\partial \rho }{\partial t}=-\frac{\partial \lbrack f(x)\rho ]}{%
\partial x}-\frac{\partial \lbrack \eta (x,t)\rho ^{q/2}]}{\partial x}+D%
\frac{\partial ^{2}\rho ^{q}}{\partial x^{2}}  \label{eq41}
\end{equation}

Now follow the procedure used earlier. It is straightforward to show that
the noise

\begin{equation}  \label{eq42}
\zeta \left( {x,t} \right) = \sum\limits_{i} {\frac{{\partial} }{{\partial x}%
}\left( {\xi _{i} \left( {t} \right)\rho ^{\nu} \left( {x,t} \right)\rho
_{i} \left( {x,t} \right)} \right)}
\end{equation}

\noindent has the same statistical property as the global noise

\begin{equation}  \label{eq43}
\zeta ^{\prime}\left( {x,t} \right) = \frac{{\partial \left[ {\eta \left( {%
x,t} \right)\rho ^{\nu + 1/2}\left( {x,t} \right)} \right]}}{{\partial x}}
\end{equation}

Where the noise $\eta $ is defined as before through equation (\ref{eq24}).

So once again we have a closed Langevin equation for the evolution of the
density function, {\it \~{n}}:

\begin{equation}  \label{eq44}
\frac{{\partial \rho} }{{\partial t}} = - \frac{{\partial \left[ {f\left( {x}
\right)\rho} \right]}}{{\partial x}} - \frac{{\partial \left[ {\eta \left( {%
x,t} \right)\rho ^{q/2}} \right]}}{{\partial x}} + D\frac{{\partial ^{2}\rho
^{q}}}{{\partial x^{2}}}
\end{equation}

Forming the average quantities yields equation (\ref{eq38}) as expected.
More to the point this equation also follows from the following coarse
grained free energy defined below together with the procedure defined by
equation (\ref{eq27}) (with {\it \^{i}} replaced by {\it \ae })

\begin{equation}  \label{eq45}
F\left[ {\rho} \right] = \int {dxV} \left( {x} \right)\rho \left( {x}
\right) + \frac{{1}}{{1 - q}}\left( {1 - \int {dx\rho ^{q}\left( {x,t}
\right)}} \right)
\end{equation}

Thus we have now a free energy functional for our system where the role of
entropy previously played by a Boltzmann expression is replaced by the
Tsallis formula.

In approaching these complex systems via the Langevin equation we can begin
to see the nature of the approximations involved. Already we have noted that
the local noise could be characterised by a different form including, for
example, correlations with agents of greater wealth. In the case of a liquid
we might consider extending the influence to include molecules in the wider
neighbourhood. For example we might replace $\rho ^{\nu} \left( {x} \right)$%
with say $\int_{ - \infty} ^{\infty} {dx^{\prime}} e^{ - \alpha |x -
x^{\prime}|}\rho ^{\nu} \left( {x^{\prime}} \right)$where \'{a} is a
parameter characterising the range of influence. Clearly \'{a} = 0 yields
the Boltzmann limit whereas \'{a} >? is consistent with the Tsallis limit.
Similarly for socio economic systems one plausible approximation could be to
replace $\rho ^{\nu} \left( {x} \right)$ with$\int_{x}^{\infty} {dx^{\prime}}
e^{ - \alpha \left( {x - x^{\prime}} \right)}\rho ^{\nu} \left( {x^{\prime}}
\right)$which assumes a peer pressure where only `wealthier' people have an
influence on individual behaviour. The reader may think of other variants.
The point we make is to suggest that the Tsallis expression is not a unique
expression. Rather it may be an approximation albeit a rather clever one
that accounts in a subtle manner for important correlations in these complex
systems.

\section*{Wealth distributions}

In a recent paper Borgas \footnote{%
E P Borges arXiv: cond-mat/0205520 v2 24 May 2002} studied data for wealth
distributions in the USA (2002), Brazil (1996), Germany (1998) and the UK
(1998). The cumulative distributions, he found could be fitted well with a q
q' Gaussian obeying the differential equation:

\begin{equation}  \label{eq46}
\frac{{1}}{{2x}}\frac{{dP}}{{dx}} = - \left( {\beta - \beta ^{\prime}}
\right)P^{q} - \beta ^{\prime}P^{q^{\prime}};\quad \quad 1 \le q^{^{\prime}}
\le q\quad and\quad 0 \le \beta ^{\prime}\le \beta
\end{equation}

Full details are given in Borgas' paper. Here we note the values he obtained
for the exponents q and q':

\begin{equation}
\begin{array}{l}
\quad \quad \quad \quad \quad \quad \quad q\quad \quad \quad q^{\prime } \\ 
USA\quad \quad \quad \quad \quad 3.8\quad \quad 1.7 \\ 
Brazil\quad \quad \quad \quad \ \ 3.5\quad \quad 2.1 \\ 
Germany\quad \quad \quad 2.9\quad \quad 1.0 \\ 
UK\quad \quad \quad \quad \quad \ \ \ 3.1\quad \ \ \ 1.0
\end{array}
\label{eq47}
\end{equation}

Empirical data suggests that q' remains constant whereas over the past 30
years there has been a small increase in the values for q.

Return to equation (\ref{eq40}). Suppose the force acting on agent {\it i}
arises from not only pair forces but also higher order many body forces. In
this case it can be expressed in terms of a many body expansion,

\begin{equation}  \label{eq48}
f\left( {x,\left[ {\rho} \right]} \right) = a + b\rho + c\rho ^{2} + ....
\end{equation}

And a is identified with external effects; b with one body interactions, c
with 2 body interactions, etc. If we invoke the mechanism used in the Sznajd
model, then we assume agents do not respond to single body forces. Only two
body forces affect the agent dynamics. So we must set b=0. Now examine the
stationary solution of equation (\ref{eq40}):

\QTP{Body Math}
\begin{equation}
\frac{{dP^{q}}}{{dx}}=aP+cP^{3}  \label{eq49}
\end{equation}

If q $\sim $ 1 then the exponents are close to those of the European
countries, UK and Germany. If q < 1 we may write the term on the LHS of
equation (\ref{eq49}) as $qP^{q - 1}{{dP} 
\mathord{\left/ {\vphantom {{dP} {dx}}} 
\right. \kern-\nulldelimiterspace} {dx}}$ and obtain

\begin{equation}  \label{eq50}
\frac{{dP}}{{dx}} = a^{\prime}P^{2 - q} + c^{\prime}P^{4 - q}
\end{equation}

A value of q $\sim $ 0.3 now yields exponents of 1.7 and 3.7 which within
experimental error seem to be compatible with the results for the US. More
detailed analysis of the raw data is required to see how close equation (\ref
{eq50}) really is for all the data including Brazil. Furthermore, the
so-called `ankle' that is evident in the very high-income data of Borgas
remains to be explained. However this approach suggests that income data is
governed by similar mechanisms to investments and trading. Furthermore
agents within the American continent, are much more responsive to local
noise and rumour than agents in Europe. Job switching to secure higher
incomes in the US is known to be more dynamic than in Europe. Certainly that
great American investor the so-called `Sage of Omaha', Warren Buffet is on
record as saying he lives in Omaha simply to avoid being influenced by
rumours in Wall St. Perhaps Europe has traditionally been less influenced by
local rumour and noise. More detailed studies might cast more light on this
feature.

\section*{References}

\bigskip 

Figure caption

\par Figure 1. \par Normalised cumulative distribution functions for the wealth in both the standard GLV model of Solomon and the LV model with peer pressure (equation
(15)). The fit for the highlighted region (in yellow) is in good agreement with the analytical (mean-field) results. The last part of the distribution displays a sharper drop-off. Finite-size effects may account for the discrepancy between analytical and numerical results. \par

\end{document}